\documentclass[conference]{IEEEtran}
\IEEEoverridecommandlockouts
\usepackage{cite}
\usepackage{amsmath,amssymb,amsfonts}
\usepackage{algorithmic}
\usepackage{algorithm}
\usepackage{graphicx}
\usepackage{svg}
\graphicspath{ {./images/} }
\usepackage{textcomp}
\usepackage{xcolor}
\def\BibTeX{{\rm B\kern-.05em{\sc i\kern-.025em b}\kern-.08em
    T\kern-.1667em\lower.7ex\hbox{E}\kern-.125emX}}
\begin{document}

\title{Optimizing Cuckoo Filter for high burst tolerance, low latency, and high throughput\\
}

\author{\IEEEauthorblockN{Aman Khalid}
\IEEEauthorblockA{\textit{Computer Science Department} \\
\textit{Saint Louis University}\\
St. Louis, USA \\
aman.khalid@slu.edu}

}

\maketitle

\begin{abstract}
In this paper, we present an implementation of a cuckoo filter for membership testing, optimized for distributed data stores operating in high workloads. In large databases, querying becomes inefficient using traditional search methods. To achieve optimal performance it is necessary to use probabilistic data structures to test the membership of a given key, at the cost of getting false positives while querying data. The widely used bloom filters can be used for this, but they have limitations like no support for deletes\cite{b1,b2}. To improve upon this we use a modified version of cuckoo filter that gives better amortized times for search, with less false positives.
\end{abstract}


\section{Introduction}
Distributed databases like Cassandra, Foundation DB, and HBase cater to the need to scale data horizontally on the cloud\cite{b4}. Depending on the size of the organization the data-store may spawn multiple clusters within data-centers. They are optimized for the unpredictable \cite{b5} nature of the cloud. Offering features like - fault tolerance through replication and high availability. Depending on the nature of the workload(high-read or high-write \cite{b6}), it is possible to optimize these databases for individual use-cases. 

Querying these databases is a challenging problem, as it requires developers to make trade-offs depending on their use-case. These decisions can have a severe impact on performance at large workloads, and therefore managing\cite{b7} throughput and latency gets difficult. 

Membership testing is a critical aspect of big data and traditional search algorithms don’t fare well at large workloads\cite{b8}. Even though it’s possible to get better performance than binary search using bloom filters, there is a way to get even better lookup performance with fewer false positives using cuckoo filters\cite{b3}. 

\subsection{Our Contributions}
\textbf{OCF: Optimized Cuckoo Filters.} Consider a standard Cuckoo Filter\cite{b3} that uses partial-key cuckoo hashing to support membership tests. As the bucket occupancy increases the number of false positives increases significantly, this can lead to an increase in average query time. Having too many misses is also an indication that the buckets in the filter are reaching capacity, which can warrant flushes in databases like Cassandra, leading to a complete rebuild of the in-memory data structures in the nodes.

To avoid this, OCF provides burst tolerance which improves latency by preventing premature flushes. It ensures proper utilization of a node’s memory by dynamically adjusting its in-memory data structures, as the items in the filter get added or deleted.

\subsection{Membership Testing in Distributed Databases}\label{AA}

There are many factors that affect how Distributed Databases are queried\cite{b11} - like the network speed, size of the data-set, and how is the data-store configured. To fulfill requests multiple sub-queries can be triggered across the data-center and the total time is the aggregate of the time spent to fetch keys at each node.

Consider sets $T, U,$ \& $V$ stored in different nodes in a data-center. We need to find Cartesian product $\displaystyle T\ \times \ U\ =\ \{( t,\ u) \ |\ t\ \in T\ \land \ u\ \in U\} \ $ \textit{s.t} $\displaystyle V_{\alpha }  >u.T\ \forall \ T \times U$. This query will first create a set of size, $s = size(T) * size(U)$ . Then it will trigger $s$ queries in $V$ to filter results in  $T \times U$. In this case, the number of look-ups on the node containing $T$ is much greater.

Even with a high fault tolerance rate, the faults per query increases exponentially, in the scenario above. Databases like Cassandra use bloom filter for query optimization, although it is possible to configure the filter in Cassandra, it does not account for sudden changes in traffic, this can lead to over or under-utilization of resources\cite{b1,b2}. Therefore having the same configurations for the Bloom filters across a cluster can lead to performance deterioration.  In this paper, we show that using OCF we can have better-amortized times\cite{b12}.

\section{Optimized Cuckoo Filter}
A limitation of the conventional bloom filters is that it does not support deletes. There are several proposals that extend the traditional Bloom Filter, but the Hash Table based approach makes it less space-efficient. Also, the number of elements to be stored must be known prior to creation. The traditional Cuckoo filter provides higher lookup performance than Bloom Filters, it also consumes less space provided the false positive rate remains below 3\% \cite{b9,b10}.

The original cuckoo filter outperforms the bloom filter in terms of memory and lookup speed. However, it fails when the maximum load goes beyond 0.9\cite{b12,b13}. There have been adaptations of the cuckoo filter in distributed databases, which suffer from this issue. We observed an occasional false negative when operating at this threshold\cite{b14}, which breaks membership testing. Therefore to run reliably in the cloud cuckoo filter needs to account for the unpredictable nature of traffic.

The design of OCF is inspired by congestion in network switches. The ability to adapt based on the extent of the load is the prime focus of our implementation\cite{b15}. OCF can be fine-tuned for different requirements. OCF can operate in two modes of operations that are selected during initialization. One is Congestion Aware (EOF) mode and the one is Primitive mode(PRE).

\subsection{Modes of Operation}
\subsubsection{Primitive}
The primitive mode(PRE) of OCF adjusts the size of the underlying filter based on static parameters. The user can choose the minimum and maximum thresholds for the size of the filter. The filter is resized when the occupancy rate reaches the threshold. Using this mode works fine for up to a million records, however, it is not advised to use PRE when the number of keys is more than one million. At that scale, subsequent deletes cause the filter to shrink linearly. However, the occupancy $O$ remains above the safe limit, which can result in false negatives and breaking the implementation.

\subsubsection{Congestion Aware}
\begin{figure}[htbp]
\centerline{\includegraphics[scale=0.25]{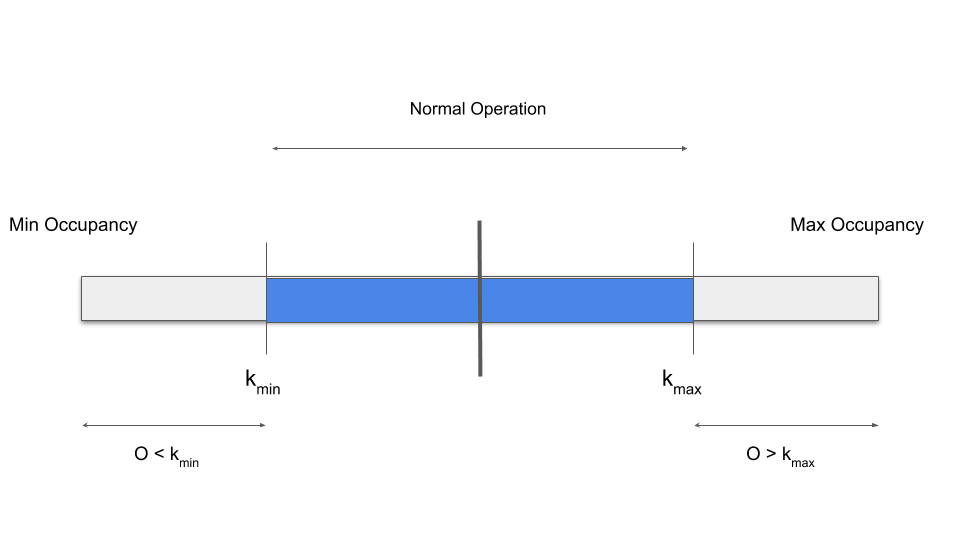}}
\caption{Visual Representation of $0 < O < 1$}
\label{window}
\end{figure}

The Congestion Aware mode (EOF) changes the filter, based on the extent to which the rate of insertion or deletion is changing in the filter. This is done by marking all the insertions and deletions beyond a value k. In Fig.~\ref{window}, the area between Min Occupancy $O_{min}$ and Max occupancy $O_{max}$ represents the value of occupancy. 

If the filter occupancy remains between the two no resize is triggered in PRE or EOF. In EOF, when $O < k_{min} \ or \ O> k_{max}$ OCF starts monitoring the changes in the filter from thereon. The new size of the filter is determined based on the rate and number of entries that get added or removed from the filter. Using this mode is safer when the number of records is more than one million as each increase or decrease takes into account the factors that caused the previous resize. It’s not recommended to use this mode for smaller workloads as PRE performs better while consuming comparatively low memory.

\subsection{OCF Parameters}
\begin{itemize}
\item \textbf{Capacity}: The capacity $c$ of the OCF filter, it’s recommended that the capacity be set twice as much as the number of elements to be inserted. This changes during run-time as the number of elements increase or decrease
\item \textbf{Bucket Size}: The size of individual buckets in the filter. The number of buckets is equal to the capacity of the filter. The recommended value for bucket size is 4 as it triggers fewer evictions while consuming less space. This parameter cannot be changed after the bucket has been created.
\item \textbf{Fingerprint Size}: Length of the fingerprints that will be stored in the buckets. Choose the size based on the total expected number of items that will be stored in the node. Choosing a lower value can cause collisions. If fingerprint size is x possible unique fingerprints are $(10)^6$.
\item \textbf{Max Displacements}: This is the number of times a filter will try to find a place to store the item. After the number of retires is reached the filter is full.
\item \textbf{Max Occupancy}: If occupancy of the filter goes above this value, the filter resets.
\item \textbf{Min Occupancy}: If occupancy of the filter reaches below this value, the filter resets.
\item \textbf{K Marker(in EOF only)}: sets the minimum and maximum threshold at which monitoring starts.
\item \textbf{Estimation Gain(in EOF only)}: Estimation Gain g Sets the rate at which growth factor $\alpha$ increases. The default value is 1/16.

\end{itemize}

\subsection{Algorithm}

In PRE mode the OCF does not account for the rate at which the filter gets filled. The occupancy of the filter is the prime factor that decides when will the filter be resized. O is calculated by Number of Items in the bucket s and Capacity c, $O = s/c$ where $0< O < 1$. When $O>O_{max}$ the bucket is doubled in size. In case when the items in bucket decrease below $O_min$ the bucket size cannot be simply reduced to half, instead the new size is calculated by $c = (c - c/10)$.
\begin{algorithm}
\begin{algorithmic}[1]
\item[1] $\textrm{When}\ O>k_{max} \ | \ O < k_{min} \  \textrm{mark the consecutive items}$
\item[2] $\textrm{Once} \ {O} \ \textrm{reaches the threshold: }$
\item[3] $ \ \ \ \textrm{Set: } \ M = (c’ * t’)/(c * t)$
\item[4] $ \ \ \ \textrm{Set: } \ \alpha = \alpha * (1-g) + g*M$
\item[5] $\textrm{\textbf{IF} } \ O<O_{max}$
\item[7] $ \ \ \ c = c - c*(1- \alpha)$
\item[8] $ \textrm{\textbf{ELSE}} $
\item[9] $ \ \ \ c = c + c*(\alpha)$

\end{algorithmic}
\caption{\label{alg:seeifrelin} Algorithm to resize bucket in EOF mode}
\end{algorithm}

The EOF mode OCF starts marking items when bucket occupancy goes beyond $k$. Once $O$ becomes greater than $O_{max}$ or less than $O_{min}$. Growth factor $\alpha$ is calculated. The value of $\alpha$ is directly proportional to $g$ and the ratio of the previous and current rates. Capacity and time before reset $c’\ \& \ t’$. Capacity and time during reset $c$ and $t$. 

\begin{table}[]
\caption{Occupancy and the Average number of false positives in EOF and PRE modes after inserting 1 million keys.}
\label{falsepos}
\begin{center}

\begin{tabular}{lrrll}
\cline{1-3}
\multicolumn{1}{|l|}{\textbf{Mode}} & \multicolumn{1}{r|}{\textbf{Occupancy}} & \multicolumn{1}{r|}{\textbf{Average False Positives}} &  &  \\ \cline{1-3}
\multicolumn{1}{|l|}{\textbf{EOF}}  & \multicolumn{1}{r|}{0.74}               & \multicolumn{1}{r|}{49}                               &  &  \\ \cline{1-3}
\multicolumn{1}{|l|}{\textbf{PRE}}  & \multicolumn{1}{r|}{0.47}               & \multicolumn{1}{r|}{32}                               &  &  \\ \cline{1-3}
                                    & \multicolumn{1}{l}{}                    & \multicolumn{1}{l}{}                                  &  & 
\end{tabular}
\end{center}
\end{table}

\section{results}
We ran our implementation on different key sizes ranging from 10000 - 1000000. We test both the modes of OCF for throughput and accuracy. Table ~\ref{falsepos} shows the comparison between EOF and PRE modes at 100000 keys. It can be seen that EOF has much higher occupancy than PRE at that size. However PRE has slightly better false positive count as it is below the 50\% occupancy. On the other hand this consumes a lot more space, and a lot of space in the filter is never utilized. 

\begin{figure}[htbp]
\centerline{\includegraphics[scale=0.35]{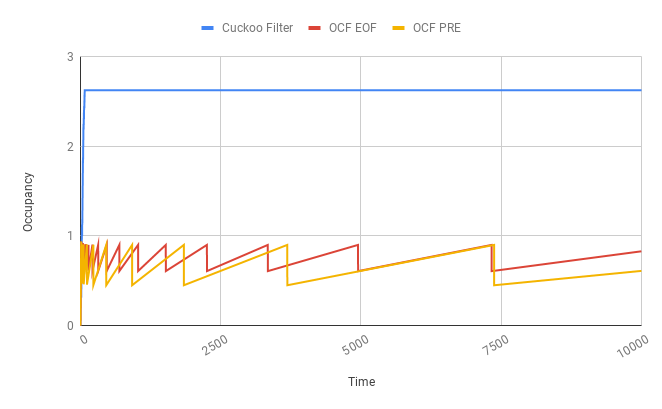}}
\caption{Throughput test of EOF, PRE and traditional cuckoo filter}
\label{zigfig}
\end{figure}

In the graph ~\ref{zigfig} it can be seen that the cuckoo filter without OCF gets completely filled within first few trials of the experiment. Both EOF and PRE perform well for the first 2500 rounds, however as the number of elements increase, it can be seen that PRE get exponentially larger therefore consuming more space than necessary. Whereas EOF maintains optimal size.

\begin{figure}[htbp]
\centerline{\includegraphics[scale=0.35]{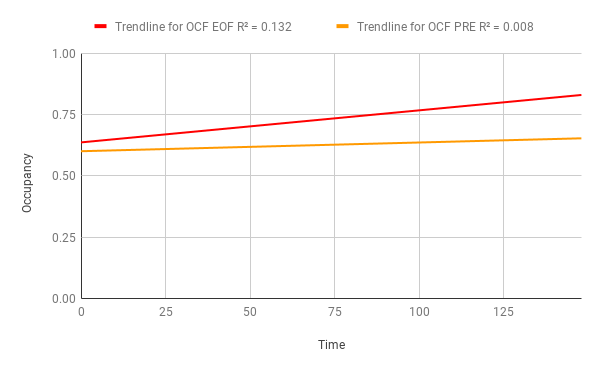}}
\caption{Treadlines of EOF and PRE}
\label{tread}
\end{figure}

It can be seen in graph ~\ref{tread}, that trendlines for EOF and OCF are similar for initial trials, as the size of the filters increase, EOF tends to maintain optimality by utilizing maximum possible space, doing this is beneficial because memory constraints become more prominent at that scale. We ran these experiments on a machine with 8 GB RAM, running intel’s i7-8750H processor with 12 cores.

\section{Conclusion}

The traditional cuckoo filter cant guarantee performance when the number of keys is larger than capacity provided. Using OCF filters we were able to extend the cuckoo filter to accept high volumes of inserts and adapt to it. Also, the traditional cuckoo filter does not have any safeguards against deleting keys that havent been inserted, trying to delete keys that were not inserted from traditional cuckoo filter removes fingerprints inserted by other keys. OCF overcomes this limitation by verifying the incoming key with the in-memory key-store, before deleting it. 
The EOF mode of the cuckoo filter saves memory and predicts the increase in traffic with reasonable accuracy, this improves as the number of trials increases, thus gives better amortized times. However PRE mode lacks this consideration and consumes almost twice as much space when number of records ~ 1000000. PRE performs marginally better false positive rates than EOF at large scale, because its filter size is twice as large.


\begin{thebibliography}{00}

\bibitem{b1} Almeida, P.S., Baquero, C., Preguiça, N.M.,\& Hutchison, D. , ``Scalable Bloom Filters,'' Inf. Process. Lett., pp. 255--261, 2007.
\bibitem{b2} Lu, Yi, and Prabhakar, Balaji and Bonomi, Flavio, Bloom filters: Design innovations and novel applications, January 2005.
\bibitem{b3} Bin Fan, Dave G. Andersen, Michael Kaminsky, and Michael D. Mitzenmacher, ``Cuckoo Filter: Practically Better Than Bloom,'' In Proceedings of the 10th ACM International on Conference on emerging Networking Experiments and Technologies (CoNEXT ’14). Association for Computing Machinery, New York, NY, USA, pp. 75--88.
\bibitem{b4} Avinash Lakshman and Prashant Malik, ``Cassandra: a decentralized structured storage system,'' SIGOPS Oper. Syst. Rev. 44, 2 (April 2010), 35--40 2010.
\bibitem{b5} D. Mittal and N. Agarwal, ``A review paper on Fault Tolerance in Cloud Computing,'' 2nd International Conference on Computing for Sustainable Global Development (INDIACom), New Delhi, 2015, pp. 31--34
\bibitem{b6} Brian F. Cooper, Adam Silberstein, Erwin Tam, Raghu Ramakrishnan, and Russell Sears, ``Benchmarking cloud serving systems with YCSB,''In Proceedings of the 1st ACM symposium on Cloud computing (SoCC ’10). Association for Computing Machinery, New York, NY, USA, 143--154, 2010.
\bibitem{b7} Boris Novikov, Natalia Vassilieva, Anna Yarygina, ``Querying Big Data,'' International Conference on Computer Systems and Technologies - CompSysTech,2012.
\bibitem{b8} Ankit R. Chadha, Rishikesh Misal, Tanaya Mokashi , ``Modified Binary Search Algorithm,'' International Journal of Applied Information Systems (IJAIS), April 2014.
\bibitem{b9} Tarkoma, Sasu, Christian Esteve Rothenberg, and Eemil Lagerspetz. "Theory and practice of bloom filters for distributed systems." IEEE Communications Surveys \& Tutorials 14.1 (2011): 131-155.
\bibitem{b10} Graf, Thomas Mueller, and Daniel Lemire. "Xor Filters: Faster and Smaller Than Bloom and Cuckoo Filters." Journal of Experimental Algorithmics (JEA) 25.1 (2020): 1-16.

\bibitem{b11} Ghaemi, Reza, et al. "Evolutionary query optimization for heterogeneous distributed database systems." World Academy of science 43 (2008): 43-49.

\bibitem{b12} Eppstein, David. "Cuckoo filter: Simplification and analysis." arXiv preprint arXiv:1604.06067 (2016).
\bibitem{b13} Fleming, Noah. “Cuckoo Hashing and Cuckoo Filters.” (2018).
\bibitem{b14} Almeida, Paulo Sérgio, et al. "Scalable bloom filters." Information Processing Letters 101.6 (2007): 255-261.

\bibitem{b15} Lang, Harald, et al. "Performance-optimal filtering: Bloom overtakes cuckoo at high throughput." Proceedings of the VLDB Endowment 12.5 (2019): 502-515.
\end{thebibliography}
\end{document}